\documentclass[12pt,english]{article}

\usepackage[T1]{fontenc}
\usepackage[latin9]{inputenc}
\usepackage{geometry}
\usepackage{amsmath}
\usepackage{amsthm}
\usepackage{setspace}
\makeatletter
\newcommand{\lyxaddress}[1]{
	\par {\raggedright #1
	\vspace{1.4em}
	\noindent\par}
}

\makeatother

\usepackage{babel}
\begin{document}
\title{\textbf{Equivalence Principle and Machian origin of extended gravity }}
\maketitle
\begin{center}
\textbf{\large{}$^{1}$Elmo Benedetto, $^{2,*}$Christian Corda and
$^{3}$Ignazio Licata}{\large\par}
\par\end{center}

\lyxaddress{\textbf{\large{}$^{1}$}\textbf{Department of Computer Science, University
of Salerno, Via Giovanni Paolo II, 132, 84084 Fisciano (Sa), Italy,
E-mail: }\textbf{\emph{ebenedetto@unisa.it }}\textbf{; }\textbf{\large{}$^{2}$}\textbf{SUNY
Polytechnic Institute, 13502 Utica, New York, USA and Università eCampus,
22060 Novedrate (CO), Lombardy, Italy, E-mail: }\textbf{\emph{cordac.galilei@gmail.com,
$^{*}$Corresponding Author;}}\textbf{ }\textbf{\large{}$^{3}$}\textbf{Institute
for Scientific Methodology (ISEM) Palermo, Italy, E-mail: }\textbf{\emph{ignazio.licata3@gmail.com}}\textbf{.}}
\begin{abstract}
Chae's analyses on GAIA observations of wide binary stars have fortified
the paradigm of extended gravity with particular attention to MOND-like
theories. We recall that, starting from the origin of Einstein's general
relativity, the request of Mach on the structure of the theory has
been the core of the foundational debate. This issue is strictly connected
with the nature of the mass-energy equivalence. This was exactly the
key point that Einstein used to derive the same general relativity.
On the other hand, the current requirements of particle physics and
the open questions within extended gravity theories, which have recently
been further strengthened by analyses of GAIA observations, request
a better understanding of the Equivalence Principle. By considering
a direct coupling between the Ricci curvature scalar and the matter
Lagrangian a non geodesic ratio between the inertial and the gravitational
mass can be fixed and MOND-like theories are retrieved at low energies.
\end{abstract}
\begin{quote}
\textbf{\emph{Essay written for the Gravity Research Foundation 2024
Awards for Essays on Gravitation}}
\end{quote}
\emph{The Science of Mechanics} by Ernst Mach \cite{key-1} had a
strong influence on Einstein and was very important in the development
of general relativity. In Newtonian theory, acceleration is absolute.
Newton deduced the existence of an absolute rotation in the famous
gedankenexperiment of the rotating bucket filled with water, by observing
the curved surfaces on the water. In that way, the inertia was explained
via a sort of resistance to motion in the absolute space which, in
turn, comes to be an agent and not a mere physical theater of coordinates,
although unspecified. The philosopher George Berkeley, in his De Motu
(1721), was the first who questioned the reasoning of Newton. He can
be considered the precursor of Mach and Einstein \cite{key-2}. In
fact, after more than 150 years, Mach strongly criticized Newton\textquoteright s
absolute space by concluding that the inertia should be an interaction
which requires other bodies to manifest itself. Thus, it would make
no sense in a Universe consisting of just a single mass. Mach's approach
proposes a total relational symmetry and every motion, uniform or
accelerated, makes sense only in reference to other bodies. Hence,
the so called \emph{Mach Principle} implies that the inertia of a
body is not an intrinsic property, but depends on the mass distribution
in the rest of the Universe instead. Although Einstein was very fascinated
by Mach reasoning, Mach Principle is not fully incorporated into general
relativity's field equations \cite{key-3}. The challenge of a\emph{
Machian gravitational physics} was accepted several times (though
less than expected) in the context of both classical and quantum theories.
An example is Narlikar\textquoteright s theory with variable mass,
which was derived from Wheeler-Feynman-like action at a distance theory
\cite{key-4,key-5}. Sciama's theory \cite{key-6} sees the inertia
as \textquotedblleft gravitational closeness\textquotedblright{} (and
the perfect equivalence) under the precise cosmological condition
$G\rho\frac{r^{2}}{c^{2}}=1,$ where $r$ is the radius of the universe,
$\rho$ the density, $c$ is the speed of light and $G$ the Newtonian
gravitational constant. In a quantum framework and in Higgs times,
the problem becomes more complex {[}7\textendash 11{]}. 
\begin{quote}
Einstein often stressed that some Machian effects should be present
in general relativity. In particular, in the famous Lectures of 1921
\cite{key-12} he argued that in general relativity there are the
following effects:
\end{quote}
\begin{enumerate}
\item \emph{The inertia of a body must increase when ponderable masses are
piled up in its neighbourhood.}
\item \emph{A body must experience an accelerating force when neighboring
masses are accelerated and the force must be in the same direction
as that acceleration. }
\item \emph{A rotating hollow body must generate inside of itself a Coriolis
field which deflects moving bodies in the sense of the rotation and
a radial centrifugal field as well.}
\end{enumerate}
Following Einstein's reasoning one considers the geodesic equation
\begin{equation}
\frac{d^{2}x_{\mu}}{ds^{2}}+\varGamma_{\mu}^{\alpha\beta}\frac{dx_{\alpha}}{ds}\frac{dx_{\beta}}{ds}=0.\label{eq: geodesic equation}
\end{equation}
In the weak-field approximation, Einstein found a metric, representing
the gravitational field due to a distribution of small masses corresponding
to a density $\sigma$ and having small velocities $\frac{dx^{i}}{ds}$,
which is 
\begin{equation}
\begin{array}{c}
g_{00}=1-\frac{2G}{^{c^{2}}}\int\frac{\sigma dV}{r}\\
\\
g_{0i}=\frac{4G}{^{c^{2}}}\int\frac{dx^{i}}{ds}\frac{\sigma dV}{r}\\
\\
g_{ij}=-\delta_{ij}\left(1+\frac{2G}{^{c^{2}}}\int\frac{\sigma dV}{r}\right).
\end{array}\label{eq: Einstein metric}
\end{equation}
Then, defining 
\begin{equation}
\begin{array}{c}
\overline{\sigma}\equiv\frac{G}{^{c^{2}}}\int\frac{\sigma dV}{r}\\
\\
A\equiv\frac{4G}{^{c^{2}}}\int\frac{\sigma vdV}{r}
\end{array}\label{eq: defining}
\end{equation}
one finds the equation of motion as 
\begin{equation}
\frac{d}{dx^{0}}\left[\left(1+\overline{\sigma}\right)v\right]=\nabla\overline{\sigma}+\frac{dA}{dx^{0}}+\left(\nabla\land A\right)\land v.\label{eq: equation of motion}
\end{equation}
Einstein's interpretation was that the inertial mass $m_{i}$ is proportional
to $\left(1+\overline{\sigma}\right)$ and, consequently, it should
increase when ponderable masses approach the test body
\begin{equation}
m_{i}=m_{g}\left(1+\frac{G}{^{c^{2}}}\int\frac{\overline{\sigma}dV}{r}\right),\label{eq: increase}
\end{equation}
where $m_{g}$ is the gravitational mass. Brans' interpretation \cite{key-13},
accepted by several physicists, was that only the second and third
effect are contained in general relativity. At first glance it would
seem that, if Einstein's interpretation were correct, there would
be a violation of the Equivalence Principle. However, it should be
emphasized that all bodies with different inertial masses are still
falling with the same acceleration in a gravitational field. Darabi
\cite{key-14} analyzed what he called \emph{Modified Mach Principle}
in the context of an expanding universe. He suggested the following
definitions for the inertial mass within and beyond the bulge of galaxies
as 
\begin{equation}
\begin{array}{ccc}
m_{i}=C &  & r\leq R_{0}\\
\\
m_{i}=\frac{C'}{r}=m_{g}\frac{R_{0}}{r} &  & r>R_{0},
\end{array}\label{eq: 3 per 3}
\end{equation}
where $R_{0}$ is the size of the bulge and $C$ and $C'$ are constants:
the first one is inertial mass versus gravitational interaction within
the bulge, and the second one is inertial mass versus cosmological
expansion beyond the bulge. Then, the introduction of a genuine Mach's
principle seems to have a need for re-introduction of the distinction
between inertial mass and gravitational mass, hidden under the metric
of general relativity and the strong form of the Equivalence Principle,
which locally turns out to be always valid in support of the structure
of general relativity, both from the classical \cite{key-15} and
quantum \cite{key-16} point of view. On the other hand, the equivalence
between inertial and gravitational mass is the axiomatic and constructive
keystone not only of general relativity, but of all the metric theories
of gravity. One is then faced with the foundational problem of the
interpretation of the formalism able to establish the equivalence
principle on the physical meaning of the relationship between inertial
and gravitational mass. This, could be connected with another foundational
problem in cosmology and gravitation, the one concerning the nature
of Dark Matter, which is one of the unsolved mysteries in Science
since C. Zwicky measured the velocity dispersion of the Coma cluster
of galaxies \cite{key-17}. Let us consider the equation 
\begin{equation}
m_{i}\frac{v^{2}}{r}=\frac{GM_{g}m_{g}}{r^{2}},\label{eq: moto circolare}
\end{equation}
where $m_{i}$ is a body that rotates around a gravitational mass
$M_{g}$ over a constant radius $r.$ It is well known that the famous
Milgrom's relation, which is founded on MOND \cite{key-18,key-19},
\begin{equation}
v=\sqrt[4]{GM_{g}a_{0}},\label{eq: MOND velocity}
\end{equation}
 with $a_{0}\approx10^{-10}\:\frac{m}{s^{2}},$ is in agreement with
various observational evidences, although not with all, and has been
recently endorsed by Chae's analyses on GAIA observations of wide
binary stars \cite{key-20,key-21}. Hence, by combining Eqs. (\ref{eq: moto circolare})
and (\ref{eq: MOND velocity}) one writes 
\begin{equation}
v^{2}=\frac{GM_{g}}{r}\frac{m_{g}}{m_{i}}=\sqrt{GM_{g}a_{0}},\label{eq: combiniamo}
\end{equation}
 which imples that Milgrom's acceleration $a_{0}$ depends on the
ratio between gravitational and inertial mass as 
\begin{equation}
a_{0}=\left(\frac{m_{g}}{m_{i}}\right)^{2}\frac{GM_{g}}{r^{2}}.\label{eq: Milgrom's acceleration}
\end{equation}
In other words, MOND dynamics could depend on violations of the Equivalence
Principle at large distances. This is not in contrast to today's strong
empirical evidence of the Equivalence Principle \cite{key-22}, as
observations and experiments on the equivalence between inertial mass
and gravitational mass are conducted on Earth, or at least within
the Solar System. Let us see the situation in another way. From Eq.
(\ref{eq: combiniamo}) one also gets 
\begin{equation}
\frac{m_{g}}{m_{i}}=\sqrt{\frac{a_{0}r^{2}}{GM_{g}}.}\label{eq: rapporto masse}
\end{equation}
Rather than interpreting $a_{0}$ from the kinematic point of view
one can interpret it in terms of a gravitational field by writing
\begin{equation}
\frac{m_{g}}{m_{i}}=\sqrt{\frac{g_{0}}{g},}\label{eq: rapporto accelerazioni}
\end{equation}
where $g=\frac{GM_{g}}{r^{2}}$ is the standard Newtonian acceleration.
According to Mach's interpretation, the inertial mass of a body arises
as a consequence of its interactions with the Universe. Thus, one
assumes that 
\begin{equation}
\frac{m_{g}}{m_{i}}\equiv\mu,\label{eq: mu}
\end{equation}
where $\mu=1$ for $\left|\frac{g_{0}}{g}\right|\ll1$ (relatively
small distances) and $\mu=\sqrt{\frac{g_{0}}{g}}$ for $\left|\frac{g_{0}}{g}\right|\gg1$
(large distances). A possible form of $\mu$ could be {[}25\textendash 27{]}
\begin{equation}
\mu\equiv\sqrt{\frac{g_{0}+g}{g}},\label{eq: definizione mu}
\end{equation}
where in this case $g_{0}$ is the Machian gravitational field generated
by all the masses of the Universe different from $M_{g}$. It can
easily be verified that, when $g\gg g_{0}$ the circular velocity
decreases with increasing distance from $M_{g}$, according to the
Newtonian law, but, when $g\ll g_{0}$ one obtains 
\begin{equation}
v^{2}=\frac{GM_{g}}{r}\sqrt{\frac{g_{0}}{g}}=GM_{g}\sqrt{\frac{g_{0}}{GM_{g}}}=\sqrt{GM_{g}g_{0}},\label{eq: ottiene}
\end{equation}
which leads immediately to 
\begin{equation}
v=\sqrt[4]{GM_{g}g_{0}}.\label{eq: velocit=0000E0 breakdoun PE}
\end{equation}
Obviously, the value of $g_{0}$ which fits the majority of the data
of galaxies rotation curves is about $10^{-10}\:\frac{m}{s^{2}}.$
If, on the one hand, the relations (\ref{eq: MOND velocity}) and
(\ref{eq: velocit=0000E0 breakdoun PE}) coincide from the mathematical
point of view, on the other hand from the physical point of view the
situation is different. At every point in the Universe Newton second
law continues to be valid even in the presence of small accelerations.
This is due to the fact that the Machian gravitational field generated
by all the masses of the Universe different from $M_{g}$, which still
has Newtonian origin, dominates over the Newtonian gravitational field
generated by $M_{g}$. It is important to ask what could be the geometric-relativistic
counterpart of the weak field approach developed so far. An intriguing
interpretation in a geometric-relativistic sense is the following.
In 2007 Bertolami and others \cite{key-23} proposed an explicit coupling
between an arbitrary function of the scalar curvature, $R,$ and the
Lagrangian density of matter in the framework of $f(R)$ gravity via
the action 
\begin{equation}
S=\int\left\{ \frac{1}{2\kappa}f_{1}(R)+\left[1+\lambda f_{2}(R)\right]L_{m}\right\} \sqrt{-g}dx^{4},\label{eq: azione estesa}
\end{equation}
where $\kappa\equiv8\pi Gc^{-4}$ is the Einstein gravitational constant
and $L_{m}$ is the Lagrangian density corresponding to matter. By
setting $f_{1}(R)=f_{2}(R)=R,$ $\lambda\ll\frac{1}{2\kappa},$ then
the theory arising from the corresponding action 
\begin{equation}
S=\int\left(\frac{1}{2\kappa}R+\lambda RL_{m}+L_{m}\right)\sqrt{-g}dx^{4},\label{eq: azione weak deviation GR}
\end{equation}
which only includes a weak coupling between the Ricci scalar and the
matter Lagrangian, represents a weak deviation from standard general
relativity and can, in principle, pass the solar system terms. Adapting
the analysys in \cite{key-23} to the theory arising from the action
of Eq. (\ref{eq: azione weak deviation GR}), one introduces the standard
energy-momentum tensor of a perfect fluid $T_{\mu\nu}^{(m)}\equiv\left(\epsilon+p\right)u_{\mu}u_{\nu}$,
where $\epsilon$ and $p$ are the overall energy density and the
pressure, respectively. $u_{\mu}$ is the four-velocity satisfying
$u_{\mu}u^{\mu}=1$ and $u^{\mu}u_{\mu;\nu}=0.$ Then, one finds that
the coupling between the Ricci scalar and the matter Lagrangian generates
a non-geodesic equation compatible with a violation of the Equivalence
Principle at large distances \cite{key-23}
\begin{equation}
\frac{d^{2}x_{\mu}}{ds^{2}}+\varGamma_{\mu}^{\alpha\beta}\frac{dx_{\alpha}}{ds}\frac{dx_{\beta}}{ds}=F^{\alpha},\label{eq: non-geodesic equation}
\end{equation}
due to the presence of extra force orthogonal to the four-velocity
of the particle \cite{key-23}
\begin{equation}
F^{\alpha}=\frac{1}{\epsilon+p}\left[\frac{\lambda}{1+\lambda R}\left(L_{m}+p\right)\nabla_{\beta}R+\nabla_{\beta}p\right]h^{\alpha\beta},\label{eq: extra force}
\end{equation}
where the projection operator $h_{\mu\nu}\equiv g_{\mu\nu}-u_{\mu}u_{\nu}$
has been introduced, which satisfies $h_{\mu\nu}u^{\mu}=0.$ The weak
field limit in three dimensions of Eq. (\ref{eq: non-geodesic equation})
is \cite{key-23}
\begin{equation}
\overrightarrow{a}_{tot}=\overrightarrow{g}+\overrightarrow{a}_{ex}.\label{eq: accelerazione totale}
\end{equation}
Hence, the total acceleration $\overrightarrow{a}_{tot}$ turns out
to be the sum of the standard Newtonian one, $\overrightarrow{g},$
plus that (per unit mass) due to the presence of the extra force,
$\overrightarrow{a}_{ex}$. From Eq. (\ref{eq: accelerazione totale}),
a bit of three-dimensional geometry \cite{key-23} permits one to
write the Newtonian acceleration as 
\begin{equation}
\overrightarrow{g}=\frac{1}{2}\left(a_{tot}^{2}-g^{2}-a_{ex}^{2}\right)\frac{\overrightarrow{a}_{tot}}{a_{tot}a_{ex}}.\label{eq: accelerazione Newtoniana}
\end{equation}
In the limit in which $\overrightarrow{a}_{ex}$ dominates, that is
$g\ll a_{tot},$ one obtains \cite{key-23}
\begin{equation}
\overrightarrow{g}\simeq\frac{a_{tot}\overrightarrow{a}_{tot}}{2a_{ex}}\left(1-\frac{a_{ex}^{2}}{a_{tot}^{2}}\right)=\frac{a_{tot}}{g_{0}}\overrightarrow{a}_{tot},\label{eq: Newtoniana circa}
\end{equation}
where \cite{key-23}
\begin{equation}
a_{0}=g_{0}\equiv2a_{ex}\left(1-\frac{a_{ex}^{2}}{a_{tot}^{2}}\right)^{-1}.\label{eq: real extra force}
\end{equation}
Eq. (\ref{eq: Newtoniana circa}) implies $a_{tot}\simeq\sqrt{g_{0}g,}$
which is completely consistent with Eq. (\ref{eq: ottiene}). This
consistence enables one to combine Eq. (\ref{eq: real extra force})
with Eqs. (\ref{eq: rapporto masse}) and (\ref{eq: rapporto accelerazioni})
obtaining 
\begin{equation}
\frac{m_{g}}{m_{i}}=\sqrt{\frac{g_{0}}{g}}=\sqrt{\frac{2a_{ex}}{g\left(1-\frac{a_{ex}^{2}}{a_{tot}^{2}}\right)}}.\label{eq: rapporto masse finale}
\end{equation}
Thus, in the current approach the ratio between gravitational and
inertial mass is explained in an elegant, geometric way, via a direct
coupling between the Ricci curvature scalar and the matter Lagrangian
which generates a non geodesic motion of test particles. It should
be emphasized that, in the current approach, slightly different from
a pure MOND approach, the Machian gravitational field $g_{0}$ is
not strictly constant as it depends both on local characteristics
of the curvature and the direct coupling between curvature and matter.
This seems consistent with the fact that, although MOND appears to
be able to explain many astrophysical observations, for example the
GAIA data analyzed by Chae \cite{key-20,key-21}, it cannot explain
all of them. For example MOND does not seem completely consistent
with recent data on the rotation curve of the Milky Way because the
decreasing behavior of the rotation curve beyond 20 kpc \cite{key-24}.

In summary, in this Essay it has been shown that an approach to gravitation
conforming to the Mach Principle allows one to rediscover MOND-like
theories through a violation of the Equivalence Principle at large
distances. The geometric-relativistic counterpart of this approach
is based on a weak modification to the standard Einstein-Hilbert action
which admits a weak direct coupling between the Ricci scalar and the
Lagrangian of matter. If on the one hand the weakness of this modification
to standard general relativity allows the theory to pass, in principle,
the solar system tests, on the other hand it is precisely this direct
coupling between the Ricci scalar and the Lagrangian of matter that
generates the violation of the Equivalence Principle at large distances,
which allows one to find the MOND-like behavior in the weak field
approximation. Thus, the Machian approach in this Essay has the potential
to obtain strong observational consistency with the GAIA data analyzed
by Chae \cite{key-20,key-21}, while the non-strict constance of the
Machian gravitational field $g_{0}$ can, in principle justify variations
from the pure MOND regime in some astrophysical observations, like
the decreasing behavior of the rotation curve of the Milky Way beyond
20 kpc \cite{key-24}.

\subsubsection*{Acknowledgments}

The Authors thank Aharon Davidson, Moti Milgrom, Kenath Arun, Fabrizio
Tamburini, Mauro Carfora and Andy Beckwith for having carefully read
this Essay and for having made useful observations, criticisms and
suggestions. MariaVita Licata must be thanked for checking the English
language. The Authors also thank the Judges of the Gravity Research
Foundation for their good judgment on this Essay and an anonymous
Referee for useful comments.

The Authors dedicate this Essay to the memory of their beloved mothers.

\end{document}